\documentclass[runningheads]{llncs}
\usepackage{graphicx}
\usepackage{amsmath}
\usepackage{amsfonts}
\begin{document}
\title{COVID-19 Infection Segmentation from Chest CT Images Based on Scale Uncertainty}
\titlerunning{COVID-19 Infection Segmentation Based on Scale Uncertainty}
%
\author{Masahiro Oda\inst{1,2} \and
Tong Zheng\inst{2} \and Yuichiro Hayashi\inst{2} \and Yoshito Otake\inst{3,4} \and \\ Masahiro Hashimoto\inst{5} \and Toshiaki Akashi\inst{6} \and Shigeki Aoki\inst{6} \and Kensaku Mori\inst{2,1,4}}
\authorrunning{Masahiro Oda et al.}
%
\institute{
Information and Communications, Nagoya University, \\ Furo-cho, Chikusa-ku, Nagoya, Aichi 4648601, Japan \\
\email{moda@i.nagoya-u.ac.jp}
\and
Graduate School of Informatics, Nagoya University, Aichi, Japan \and
Graduate School of Science and Technology, Nara Institute of Science and Technology, Nara, Japan \and
Research Center for Medical Bigdata, National Institute of Informatics, Tokyo, Japan \and
Department of Radiology, Keio University School of Medicine, Tokyo, Japan \and
Department of Radiology, Juntendo University, Tokyo, Japan
 }
\maketitle              
\begin{abstract}
This paper proposes a segmentation method of infection regions in the lung from CT volumes of COVID-19 patients.
COVID-19 spread worldwide, causing many infected patients and deaths.
CT image-based diagnosis of COVID-19 can provide quick and accurate diagnosis results.
An automated segmentation method of infection regions in the lung provides a quantitative criterion for diagnosis.
Previous methods employ whole 2D image or 3D volume-based processes.
Infection regions have a considerable variation in their sizes.
Such processes easily miss small infection regions.
Patch-based process is effective for segmenting small targets.
However, selecting the appropriate patch size is difficult in infection region segmentation.
We utilize {\it the scale uncertainty} among various receptive field sizes of a segmentation FCN to obtain infection regions.
The receptive field sizes can be defined as the patch size and the resolution of volumes where patches are clipped from.
This paper proposes an infection segmentation network (ISNet) that performs patch-based segmentation and a scale uncertainty-aware prediction aggregation method that refines the segmentation result.
We design ISNet to segment infection regions that have various intensity values.
ISNet has multiple encoding paths to process patch volumes normalized by multiple intensity ranges.
We collect prediction results generated by ISNets having various receptive field sizes.
Scale uncertainty among the prediction results is extracted by the prediction aggregation method.
We use an aggregation FCN to generate a refined segmentation result considering scale uncertainty among the predictions.
In our experiments using 199 chest CT volumes of COVID-19 cases, the prediction aggregation method improved the dice similarity score from 47.6\% to 62.1\%.

\end{abstract}
\section{Introduction}

Novel coronavirus disease 2019 (COVID-19) spread worldwide, causing many infected patients and deaths.
The total number of cases and deaths related to COVID-19 are more than 212 million and 4.4 million in the world~\cite{worldmeters}.
Because of the rapid increase of COVID-19 patients, medical institutions suffer from a human resources shortage.
To prevent further infection, a quick inspection method for COVID-19 infection is pressing required.
Such quick inspection enables providing appropriate treatments to patients and curbs the spread of COVID-19.
Reverse transcriptase-polymerase chain reaction (RT-PCR) testing is used as an inspection method of COVID-19 cases.
However, it takes some hours to give a diagnosis result.
Furthermore, its sensitivity is not high, ranging from 42\% to 71\%~\cite{simpson20}.
As another choice of COVID-19 cases, CT image-based diagnosis is helpful.
The sensitivity of CT image-based COVID-19 diagnosis is reported as 97\%~\cite{ai20}.
Furthermore, a CT scan takes only some minutes.
A CT image-based computer-aided diagnosis (CAD) system for COVID-19 is expected to provide a quick and accurate diagnosis to patients.
For such CAD systems, a quantitative analysis method of the lung condition is essential.
Ground-glass opacities (GGOs) and consolidations are commonly found in the lung of viral pneumonia cases, including COVID-19.
We call them {\it infection regions}.
Automatic segmentation of them is an essential component of CAD systems.

\noindent
{\bf Related Work of COVID-19 Segmentation:}
Previously, deep learning-based automatic segmentation methods of infection regions from CT volumes of COVID-19 cases were proposed~\cite{fan20,wang20,zheng20,mahmud20,yan20}.
Fan et al.~\cite{fan20} proposed an infection region segmentation method using the Inf-Net.
The Inf-Net utilizes reverse attention and edge attention to learn features to differentiate infection and other regions.
However, because they employ 2D image-based process, 3D positional information is not utilized in their segmentation method.
Other papers also employ 2D image-based process~\cite{wang20,zheng20,mahmud20}.
Yan et al.~\cite{yan20} proposed a fully convolutional network (FCN) to segment infection and normal regions in the lung.
The FCN has contrast enhancement branches to extract features of infection regions that have various intensities.
However, because contrast information of segmentation targets is not explicitly provided to the FCN, the contrast enhancement branches' contribution to improving segmentation accuracy is limited.

\noindent
{\bf Scale Uncertainty on Patch-based Process:}
Infection regions contain many small regions.
Segmentation processes hardly segment them from whole 2D slice image or 3D volume as performed in the previous methods\cite{fan20,yan20}.
To segment such small regions, a patch-based approach is practical.
Patch-based approach is commonly employed in segmentation methods from images of large data size such as 3D CT volume~\cite{holger18,oda19,kim20} or pathological images~\cite{zhou14,xu15,wang16,tokunaga19}.
The approach is advantageous to perform deep learning-based segmentation under the limitation of GPU memory size.
In patch-based approaches, patch size is an essential factor of segmentation accuracy.
The patch size defines the size of the receptive field of segmentation models.
Also, original images or volumes can be scaled before patch clipping to change the receptive field size.
In summary, (a) {\it the resolution of original volume (VRes)} and (b) {\it the size of patch (PSize)} are essential factors for the segmentation accuracy in patch-based approaches.
In a multi-organ segmentation method~\cite{holger18}, the use of a relatively large PSize resulted in the achievement of high segmentation accuracies among large organs (liver, spleen, and stomach).
However, their segmentation accuracy of small organs (artery, vein, and pancreas) was low.
Other paper~\cite{oda19} reported that the use of small PSize is effective for small organ (artery) segmentation.
VRes and PSize should be selected to patch covers the segmentation target from their results.

In infection region segmentation, selecting appropriate VRes and PSize are difficult because the sizes of infection regions are different for each region.
If we apply a segmentation process using multiple VRess and PSizes, we can obtain multiple prediction maps having variation among them.
The variation can be considered as {\it uncertainty among scales}.
The scale uncertainty represents useful information to obtain an accurate segmentation result.
Scale uncertainty-aware aggregation process of multiple prediction maps is essential for segmenting infection regions with various sizes.

\noindent
{\bf Proposed Method and Contributions:}
We present an infection region segmentation method from a chest CT volume of a COVID-19 patient.
We developed a patch-based FCN for infection region segmentation called infection segmentation network (ISNet) to perform segmentation.
Also, we propose a scale uncertainty-aware aggregation method of prediction results.
These methods enable the segmentation of infection regions of various sizes.
ISNet has multiple encoder and a single decoder style structure.
The use of the multiple encoders enables feature extraction from infection regions with a variation of CT values.
Deep supervision is employed to improve the decoder's ability to decode the prediction result from feature value.
ISNets having various receptive field sizes are trained and used to generate prediction maps from the CT volume.
The scale uncertainty-aware prediction aggregation is applied to the multiple prediction maps to generate a final segmentation result considering uncertainty among the prediction results related to the receptive fields' size.

The contributions of this paper are (1) proposal of the ISNet with multiple encoders for feature extraction from infection regions that have a variation of CT values and (2) proposal of the scale uncertainty-aware aggregation method of prediction maps that are generated by segmentation models having a various size of receptive fields.
These methods improve the segmentation accuracy of targets with significant variations in their intensity values and sizes.

\section{Method}

The proposed method segments infection regions from a chest CT volume of a COVID-19 patient.
Set of patch volumes clipped from a CT volume is provided to ISNet.
VRes and PSize define the size of the receptive field of ISNet.
Change of the receptive field size causes variation on segmentation results (scale uncertainty).
The scale uncertainty contains valuable information to refine segmentation results.
We propose a scale uncertainty-aware aggregation process of segmentation results, which ISNets segment on various VRess and PSizes.
The process generates a refined segmentation result.

\subsection{Infection Region Segmentation by ISNet}
\label{ssec:isnet}

\begin{figure}[tb]
\begin{center}
\includegraphics[width=0.9\textwidth]{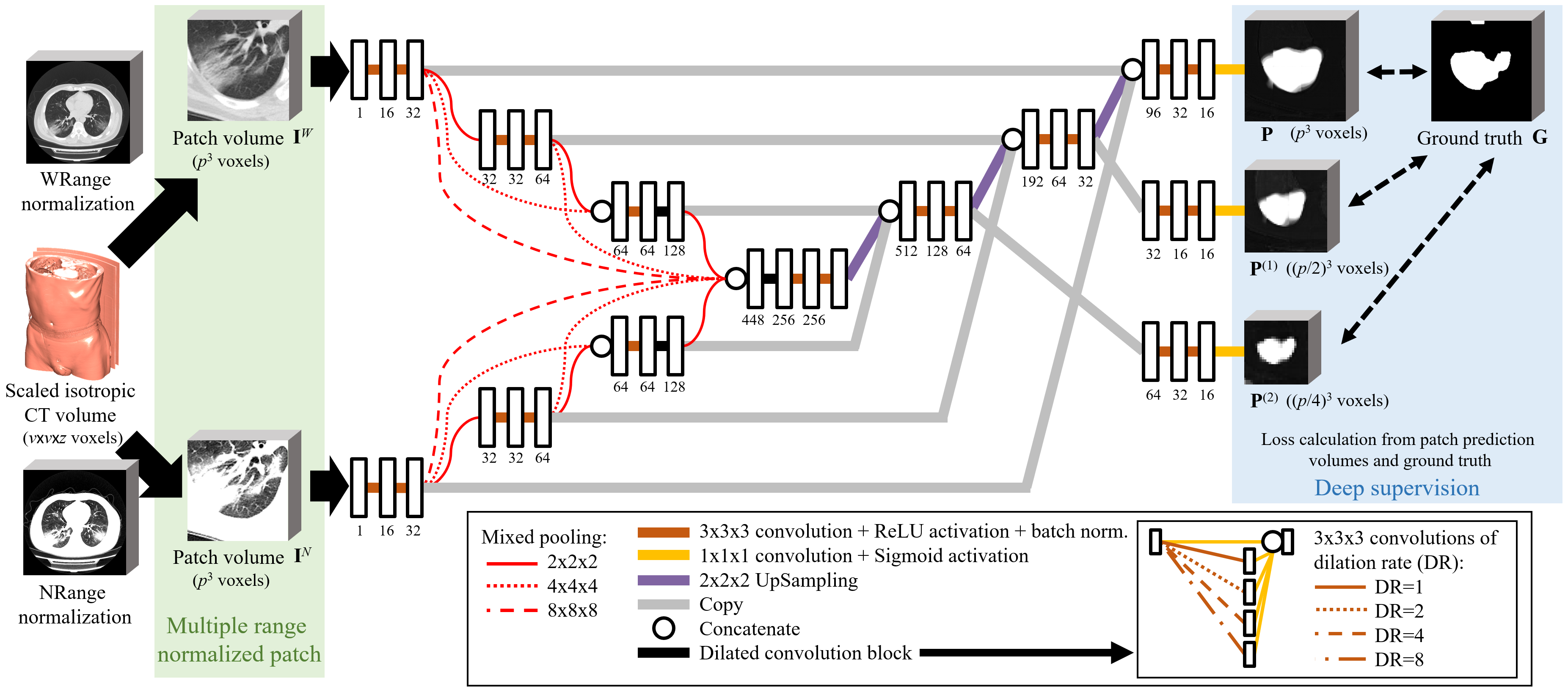}
\caption{Model structure of ISNet. Two encoders process patch volumes generated by different intensity normalizations. Encoders have dense pooling connections to bottleneck layer. Deep supervision is employed to evaluate subscale outputs.}
\label{fig:isnet}
\end{center}
\end{figure}

\noindent
{\bf Overview of Model:}
The structure of ISNet is shown in Fig.~\ref{fig:isnet}.
ISNet has multiple encoders and a single decoder.
Multiple volumes are generated from an input CT volume by applying CT value normalization by multiple value ranges to improve the segmentation accuracy of infection regions with various CT values.
Patch volumes clipped from the volumes are input to ISNet.
ISNet has multiple encoders corresponding to the multiple inputs to extract features in the CT value ranges selectively.
The encoder has dense pooling connections~\cite{playout18} that prevent loss of spatial information by pooling layers.
We employ deep supervision~\cite{zeng17,dou17} in the decoder to improve its decoding performance from feature values.

\begin{figure}[tb]
\begin{center}
\includegraphics[width=0.8\textwidth]{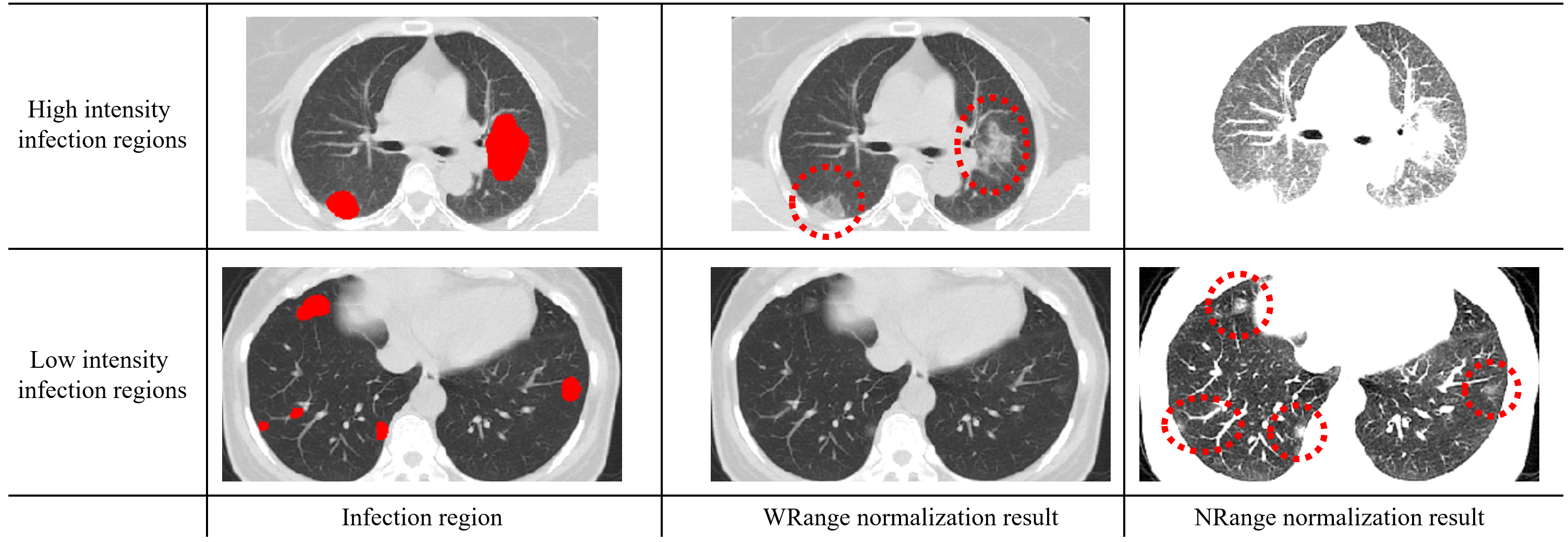}
\caption{Multiple range normalization results. Visibility of high- and low-intensity infection regions are high in WRange and NRange normalization result, respectively.}
\label{fig:ctnormalization}
\end{center}
\end{figure}

\noindent
{\bf Multiple Range Normalized Patch:}
An input CT volume is converted to a volume having an isotropic resolution in three dimensions.
Then, the volume is scaled to $v \times v \times z$ voxels maintaining the aspect ratio.
The number of voxels along the body axis $z$ differs for each CT volume depending on its scanning range.
CT values of infection regions distribute widely.
CT values of consolidations range from -300 to 100 H.U..
GGO has a lower and broader range of CT values than the consolidations, ranging from -800 to 0 H.U..
CT value normalizations by multiple ranges are adequate for such a target.
We apply CT value normalizations to the scaled CT volume using two CT value ranges, including; wide range (WRange): [-1000 H.U., 950 H.U.] and narrow range (NRange): [-1000 H.U., -400 H.U.].
Normalization results by the WRange are suitable to observe high-intensity infection regions, including consolidations and GGOs having high intensities.
Normalization results by the NRange are suitable to observe GGOs having low intensities.
Samples of normalization results are shown in Fig. \ref{fig:ctnormalization}.
Two normalized volumes, including WRange volume and NRange volume, are generated from this process.
We clip patch volumes from them at random positions.
Patch volumes clipped from the WRange and NRange volumes are described as ${\bf I}^{W}_{v,p}, {\bf I}^{N}_{v,p} \in \mathbb{R}^{p \times p \times p}$, respectively.

\noindent
{\bf Multiple Encoding Paths:}
Inputs of the ISNet are patch volumes.
We use two independent encoders to process ${\bf I}^{W}_{v,p}$ and ${\bf I}^{N}_{v,p}$.
Feature values extracted by the encoders are concatenated at the bottleneck layer.


Pooling layers are commonly used in encoder, although it reduces spatial information in feature maps.
The bottleneck layer connected after the encoder cannot receive enough spatial information.
It causes segmentation results having incorrect boundaries.
To reduce the loss of spatial information in the encoder, we adopt dense pooling connections~\cite{playout18} in the two encoders of ISNet.
The dense pooling connections provide spatial information at each resolution in the encoder to the bottleneck layer.
In the dense pooling connections, mixed poolings~\cite{playout18} are used instead of max poolings to reduce the loss of spatial information.
Furthermore, we use dilated convolution~\cite{yu16} to utilize sparsely-distributed features in convolution operations.
Dilated convolution block was implemented by connecting dilated convolutions of multiple dilation rates in parallel to obtain multiple-scales convolution results.
Some dilated convolution blocks are inserted into ISNet.

\noindent
{\bf Training:}
ISNet estimates a patch prediction volume ${\bf P}_{v,p} \in \mathbb{R}^{p \times p \times p}$ of infection regions from two input patch volumes ${\bf I}^{W}_{v,p}$ and ${\bf I}^{N}_{v,p}$.
ISNet that performs estimation from input patch volumes ($p \times p \times p$ voxels) clipped from a volume ($v \times v \times z$ voxels) can be represented as a function $f_{v,p}$.
Estimation of a patch prediction volume is formulated as
\begin{equation}
{\bf P}_{v,p} = f_{v,p}({\bf I}^{W}_{v,p}, {\bf I}^{N}_{v,p} ; \boldsymbol{\theta}_{v,p}),
\end{equation}
where $\boldsymbol{\theta}_{v,p}$ is a parameter vector for infection region segmentation.
The parameter vector is optimized in a supervised training process using CT volumes for training and their corresponding ground truth volumes ${\bf G} \in \{0,1\}^{p \times p \times p}$, whose elements 1 and 0 represent voxels in target or background regions.
We employ deep supervision~\cite{zeng17,dou17} for two subscales outputs.
Their patch prediction volumes are ${\bf P}_{v,p}^{(1)}$ and ${\bf P}_{v,p}^{(2)}$.
Their sizes are magnified to the same size as ${\bf P}_{v,p}$.
The loss function to train ISNet is defined as
\begin{equation}
L = Dice({\bf G}, {\bf P}_{v,p}) + \sum^{2}_{i=1} Dice({\bf G}, {\bf P}_{v,p}^{(i)}),
\end{equation}
where $Dice$ is the dice loss between the ground truth volume and the patch prediction volumes.

\noindent
{\bf Prediction:}
Patch volumes clipped from a CT volume for prediction are given to the trained ISNet $f_{v,p}$.
The resulting patch prediction volumes are reconstructed as the same size as the CT volume.
The reconstructed prediction volume is denoted as ${\bf R}_{v,p} \in \mathbb{R}^{v \times v \times z}$.

\subsection{Scale Uncertainty-Aware Prediction Aggregation}

The parameters $v$ and $p$ define the size of the receptive field of ISNet.
The size of the receptive field of ISNet has a relationship to its segmentation accuracy.
ISNets having various sizes of their receptive fields are trained and perform predictions, and we obtain multiple prediction volumes containing scale uncertainty from them.
We utilize the scale uncertainty-aware aggregation method of the prediction volumes.
An aggregation function is automatically trained based on each prediction volume's contribution to a segmentation result.

\begin{figure}[tb]
\begin{center}
\includegraphics[width=0.75\textwidth]{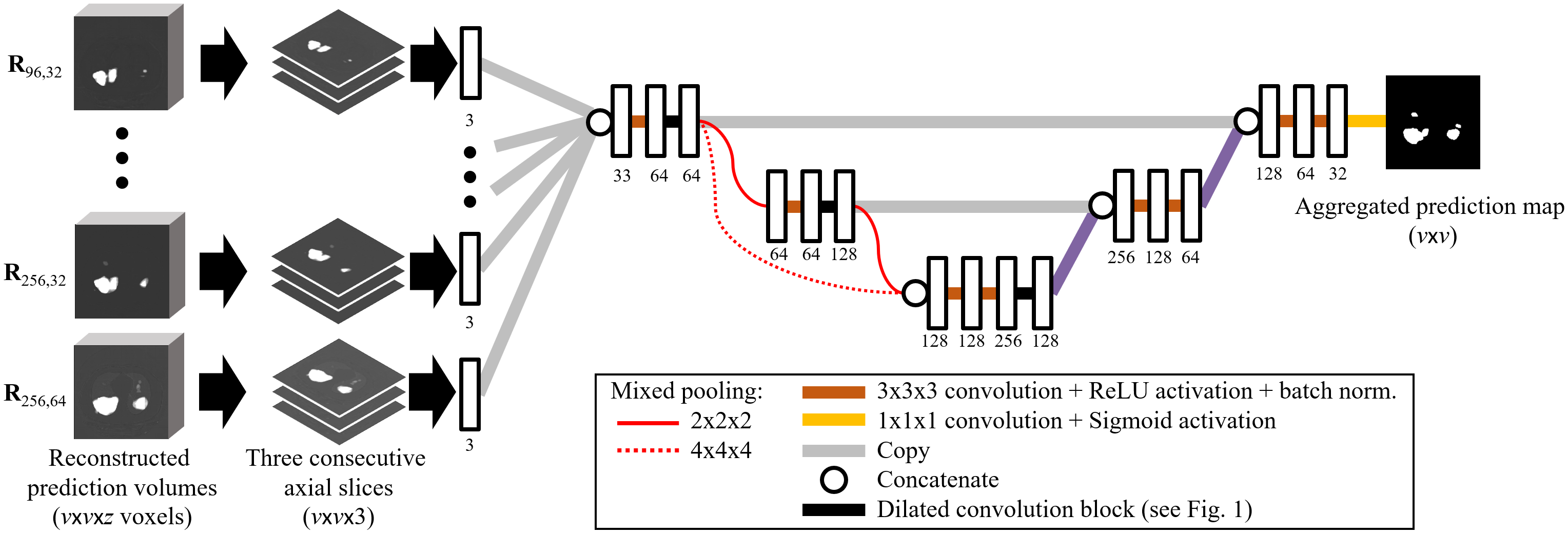}
\caption{Model structure of aggregation FCN. FCN processes axial slices obtained from prediction volumes. FCN outputs aggregation results from them.}
\label{fig:aggregationfcn}
\end{center}
\end{figure}

We train ISNets using training cases on multiple value settings of $v$ and $p$.
Using the ISNets, multiple reconstructed prediction volumes are generated from a CT volume for prediction.
We perform aggregation of them using an aggregation FCN.
The structure of the aggregation FCN is shown in Fig. \ref{fig:aggregationfcn}.
The FCN employs multiple 2D axial slice-based process.
The FCN combines given prediction results by considering variation among them and how each prediction result contributes to generating a segmentation result.
The FCN is trained using prediction volumes and their corresponding ground truth volumes of training cases.
Generalized dice loss~\cite{gdiceloss} is used to train the FCN.
In a prediction process, outputs (on 2D axial slices) of the trained aggregation FCN are reconstructed to a 3D volume that has the same resolution and size as the original CT volume and thresholded to generate a segmentation result.

\section{Experiments and Results}

We evaluated segmentation accuracy of the proposed method.
We used 199 chest CT volumes of COVID-19 patients provided by the Multi-national NIH Consortium for CT AI in COVID-19~\cite{data} via the NCI TCIA public website~\cite{tcia}.
The corresponding ground truth data of infection regions were also provided.
We conducted three-fold cross-validations in our evaluations.
Averaged precision, recall, and dice score among all CT volumes are used as the evaluation criteria.
Methods were implemented using Keras 2.2.4 and TensorFlow 1.14.0.
NVIDIA Tesla V100 GPU having 32GB memory $\times$ 1 was used to train and test the methods.

\subsection{Ablation and Comparative Study of ISNet}

We confirmed the segmentation performance of infection regions using ISNet.
Techniques explained in~\ref{ssec:isnet} are used, including the deep supervision (DS) and the multiple encoding paths (ME).
We confirmed the effectiveness of the technique in an ablation study.
Dice scores of segmentation results obtained using ISNet and ISNet without DS and ME were calculated.
ISNets were trained on parameter settings of $v=192$, $p=32$, minibatch size: 16, learning rate: $10^{-5}$, and training epochs: 40.
Adam was used as the optimization algorithm.
The segmentation result was generated by applying thresholding to the prediction volume.
Also, we compared dice scores of previous methods, including 3D U-Net~\cite{3dunet} and 3D U-Net having squeeze-and-excitation (SE) blocks~\cite{seunet1,seunet2} with ISNet.
These previous methods were applied to perform patch-based processes.
The results are shown in Table~\ref{tab:isnetresult}.
ISNet had a higher dice score than the previous methods.
Also, the use of DS and ME contributed to improving the dice score.

\begin{table}[tb]
\caption{Segmentation accuracies of ISNet using deep supervision (DS) and multiple encoding paths (ME). Accuracies of previous methods are also shown.}\label{tab:isnetresult}
\begin{center}
\begin{tabular}{|l|c|c|c|}
\hline
Method & Precision (\%) & Recall (\%) & Dice (\%) \\
\hline
ISNet (proposed method) & {\bf 58.7} & 54.6 & {\bf 56.6}\\
ISNet without DS and ME & 51.1 & 58.7 & 54.6\\
\hline
3D U-Net having SE blocks~\cite{seunet1,seunet2} & 52.3 & {\bf 59.0} & 55.5\\
3D U-Net~\cite{3dunet} & 52.5 & 55.6 & 54.0\\
\hline
\end{tabular}
\end{center}
\end{table}

\subsection{Segmentation by Aggregation FCN}

\begin{figure}[tb]
\begin{center}
\includegraphics[width=0.9\textwidth]{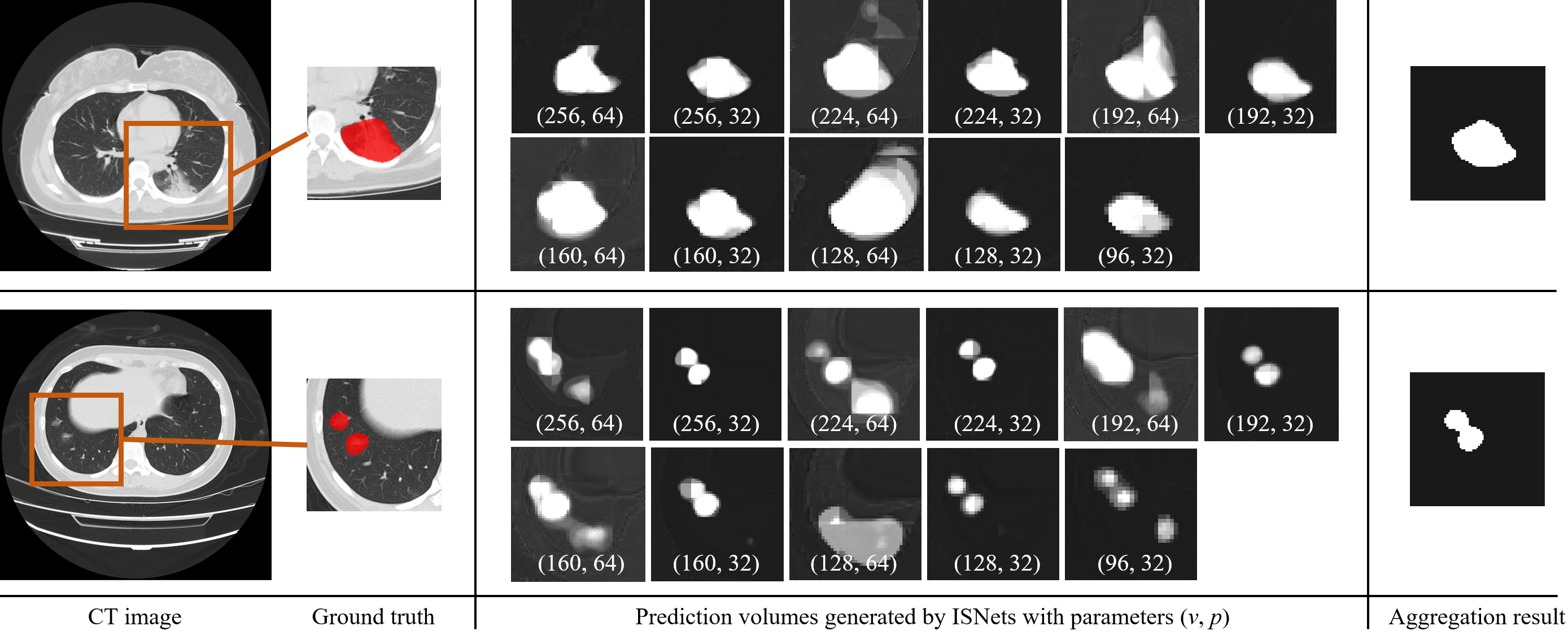}
\caption{Prediction volumes generated by ISNets with parameters $(v,p)$ and aggregation results from them. While regions in prediction volumes have considerable variation, they are adequately aggregated.}
\label{fig:result_aggregationfcn}
\end{center}
\end{figure}

We applied the scale uncertainty-aware prediction aggregation to prediction volumes generated by ISNets.
11 prediction volumes generated by ISNets trained on parameter settings $(v,p) \in \{(256,32),$ (256,64), (224,32), (224,64), (192,32), (192,64), (160,32), (160,64), (128,32), (128,64), $(96,32)\}$ were used.
The aggregation FCN was trained on parameter settings of minibatch size: 4, learning rate: $10^{-3}$, and training epochs: 5.
Adam was used as the optimization algorithm.
Generated prediction volumes and aggregation results from them are shown in Fig. \ref{fig:result_aggregationfcn}.
Segmentation accuracies of ISNets and aggregation result are shown in Table \ref{tab:aggregationresult}.
Accuracies were improved in all criteria by the aggregation.

\begin{table}[tb]
\caption{Segmentation accuracies of ISNets and aggregation result from them.}\label{tab:aggregationresult}
\begin{tabular}{|c|c|c|c|}
\hline
Method & Precision (\%) & Recall (\%) & Dice (\%) \\
\hline
Aggregation FCN & {\bf 63.7} & {\bf 60.5} & {\bf 62.1}\\
Mean $\pm$ S.D. of 11 ISNets (before aggregation) & $45.1\pm 12.3$ & $52.6 \pm 3.7$ & $47.6 \pm 8.1$\\
\hline
\end{tabular}
\end{table}

\section{Discussion and Conclusions}

Segmentation of the lung infection region is difficult because it has significant variations in its CT values and shapes.
We developed ISNet with the multiple range normalized patch processing paths and the scale uncertainty-aware prediction aggregation process to tackle infection segmentation having such difficulties.
ISNet achieved higher segmentation accuracy than the previous methods, as shown in Table~\ref{tab:isnetresult}.
Also, we confirmed the effectiveness of techniques, including deep supervision and multiple encoding paths in the ablation study.
The scale uncertainty-aware prediction aggregation process improved the dice similarity score of the segmentation result.
We used multiple prediction volumes generated by using multiple ISNets with various receptive fields' sizes.
Because segmentation abilities and effective segmentation target sizes are different among the ISNets, an appropriate aggregation process of the prediction volumes can generate an accurate segmentation result.
The aggregation FCN with trainable aggregation parameters was successfully built using training data.
The evaluation result obtained in the cross-validation proved that the trained aggregation FCN has a high generalization ability to perform segmentation from prediction volumes.

This paper proposed a segmentation method of infection regions in the lung from a CT volume of a COVID-19 patient.
To segment infection regions having variations of CT value and size, we proposed ISNet and the scale uncertainty-aware prediction aggregation process.
In our experiments, the aggregation process improved segmentation accuracy from individual ISNet results.
Future work includes increasing variety of the receptive field sizes to process in the prediction aggregation process and development of a CAD system for COVID-19 diagnosis.

\subsubsection*{Acknowledgements.} 
Parts of this research were supported by the AMED Grant Numbers 18lk1010028s0401, JP19lk1010036, JP20lk1010036, JP20lk1010025, the NICT Grant Number 222A03, the JST CREST Grant Number JPMJCR20D5, and the MEXT/JSPS KAKENHI Grant Numbers 26108006, 17H00867, 17K20099.

%
%
%
%

\end{document}